\documentstyle[twoside]{article}

\catcode`\@=11
\long\def\@makefntext#1{
\protect\noindent \hbox to 3.2pt {\hskip-.9pt  

$^{{\eightrm\@thefnmark}}$\hfil}#1\hfill}		

\def\@makefnmark{\hbox to 0pt{$^{\@thefnmark}$\hss}}	

\def\ps@myheadings{\let\@mkboth\@gobbletwo
\def\@oddhead{\hbox{}
\rightmark\hfil\eightrm\thepage}   

\def\@oddfoot{}\def\@evenhead{\eightrm\thepage\hfil
\leftmark\hbox{}}\def\@evenfoot{}
\def\sectionmark##1{}\def\subsectionmark##1{}}

\oddsidemargin=\evensidemargin
\newcounter{sectionc}\newcounter{subsectionc}\newcounter{subsubsectionc}
\renewcommand{\section}[1] {\vspace{12pt}\addtocounter{sectionc}{1} 

\setcounter{subsectionc}{0}\setcounter{subsubsectionc}{0}\noindent 

	{\tenbf\thesectionc. #1}\par\vspace{5pt}}
\renewcommand{\subsection}[1] {\vspace{12pt}\addtocounter{subsectionc}{1} 

	\setcounter{subsubsectionc}{0}\noindent 

	{\bf\thesectionc.\thesubsectionc. {\kern1pt \bfit #1}}\par\vspace{5pt}}
\renewcommand{\subsubsection}[1] {\vspace{12pt}\addtocounter{subsubsectionc}{1}
	\noindent{\tenrm\thesectionc.\thesubsectionc.\thesubsubsectionc.
	{\kern1pt \tenit #1}}\par\vspace{5pt}}
\newcommand{\nonumsection}[1] {\vspace{12pt}\noindent{\tenbf #1}
	\par\vspace{5pt}}

\newcounter{appendixc}
\newcounter{subappendixc}[appendixc]
\newcounter{subsubappendixc}[subappendixc]
\renewcommand{\thesubappendixc}{\Alph{appendixc}.\arabic{subappendixc}}
\renewcommand{\thesubsubappendixc}
	{\Alph{appendixc}.\arabic{subappendixc}.\arabic{subsubappendixc}}

\renewcommand{\appendix}[1] {\vspace{12pt}
        \refstepcounter{appendixc}
        \setcounter{figure}{0}
        \setcounter{table}{0}
        \setcounter{lemma}{0}
        \setcounter{theorem}{0}
        \setcounter{corollary}{0}
        \setcounter{definition}{0}
        \setcounter{equation}{0}
        \renewcommand{\thefigure}{\Alph{appendixc}.\arabic{figure}}
        \renewcommand{\thetable}{\Alph{appendixc}.\arabic{table}}
        \renewcommand{\theappendixc}{\Alph{appendixc}}
        \renewcommand{\thelemma}{\Alph{appendixc}.\arabic{lemma}}
        \renewcommand{\thetheorem}{\Alph{appendixc}.\arabic{theorem}}
        \renewcommand{\thedefinition}{\Alph{appendixc}.\arabic{definition}}
        \renewcommand{\thecorollary}{\Alph{appendixc}.\arabic{corollary}}
        \renewcommand{\theequation}{\Alph{appendixc}.\arabic{equation}}
        \noindent{\tenbf Appendix \theappendixc #1}\par\vspace{5pt}}
\newcommand{\subappendix}[1] {\vspace{12pt}
        \refstepcounter{subappendixc}
        \noindent{\bf Appendix \thesubappendixc. {\kern1pt \bfit #1}}
	\par\vspace{5pt}}
\newcommand{\subsubappendix}[1] {\vspace{12pt}
        \refstepcounter{subsubappendixc}
        \noindent{\rm Appendix \thesubsubappendixc. {\kern1pt \tenit #1}}
	\par\vspace{5pt}}

\newcommand{\textlineskip}{\baselineskip=13pt}
\newcommand{\smalllineskip}{\baselineskip=10pt}

\def\eightcirc{
\begin{picture}(0,0)
\put(4.4,1.8){\circle{6.5}}
\end{picture}}
\def\eightcopyright{\eightcirc\kern2.7pt\hbox{\eightrm c}}

\newcommand{\copyrightheading}[1]
	{\vspace*{-2.5cm}\smalllineskip{\flushleft
	{\footnotesize CU-TP-783}\\
	{\footnotesize hep-th/9609137}\\
	 }}

\def\abstracts#1#2#3{{
	\centering{\begin{minipage}{4.5in}\baselineskip=10pt\footnotesize
	\parindent=0pt #1\par 

	\parindent=15pt #2\par
	\parindent=15pt #3
	\end{minipage}}\par}}

\newcommand{\bibit}{\nineit}

\renewenvironment{thebibliography}[1]
	{\frenchspacing
	 \ninerm\baselineskip=11pt
	 \begin{list}{\arabic{enumi}.}
	{\usecounter{enumi}\setlength{\parsep}{0pt}
	 \setlength{\leftmargin 12.7pt}{\rightmargin 0pt} 
	 \setlength{\itemsep}{0pt} \settowidth
	{\labelwidth}{#1.}\sloppy}}{\end{list}}

\newcounter{itemlistc}
\newcounter{romanlistc}
\newcounter{alphlistc}
\newcounter{arabiclistc}

\newcommand{\fcaption}[1]{
        \refstepcounter{figure}
        \setbox\@tempboxa = \hbox{\footnotesize Fig.~\thefigure. #1}
        \ifdim \wd\@tempboxa > 5in
           {\begin{center}
        \parbox{5in}{\footnotesize\smalllineskip Fig.~\thefigure. #1}
            \end{center}}
        \else
             {\begin{center}
             {\footnotesize Fig.~\thefigure. #1}
              \end{center}}
        \fi}

\newcommand{\tcaption}[1]{
        \refstepcounter{table}
        \setbox\@tempboxa = \hbox{\footnotesize Table~\thetable. #1}
        \ifdim \wd\@tempboxa > 5in
           {\begin{center}
        \parbox{5in}{\footnotesize\smalllineskip Table~\thetable. #1}
            \end{center}}
        \else
             {\begin{center}
             {\footnotesize Table~\thetable. #1}
              \end{center}}
        \fi}

\def\@citex[#1]#2{\if@filesw\immediate\write\@auxout
	{\string\citation{#2}}\fi
\def\@citea{}\@cite{\@for\@citeb:=#2\do
	{\@citea\def\@citea{,}\@ifundefined
	{b@\@citeb}{{\bf ?}\@warning
	{Citation `\@citeb' on page \thepage \space undefined}}
	{\csname b@\@citeb\endcsname}}}{#1}}

\newif\if@cghi
\def\cite{\@cghitrue\@ifnextchar [{\@tempswatrue
	\@citex}{\@tempswafalse\@citex[]}}
\def\citelow{\@cghifalse\@ifnextchar [{\@tempswatrue
	\@citex}{\@tempswafalse\@citex[]}}
\def\@cite#1#2{{$\null^{#1}$\if@tempswa\typeout
	{IJCGA warning: optional citation argument 

	ignored: `#2'} \fi}}

\def\pmb#1{\setbox0=\hbox{#1}
	\kern-.025em\copy0\kern-\wd0
	\kern.05em\copy0\kern-\wd0
	\kern-.025em\raise.0433em\box0}

\def\fnt#1#2{\footnotetext{\kern-.3em
	{$^{\mbox{\scriptsize #1}}$}{#2}}}

\def\fpage#1{\begingroup
\voffset=.3in
\thispagestyle{empty}\begin{table}[b]\centerline{\footnotesize #1}
	\end{table}\endgroup}

\font\tenrm=cmr10
\font\tenit=cmti10 

\font\tenbf=cmbx10
\font\bfit=cmbxti10 at 10pt
\font\ninerm=cmr9
\font\nineit=cmti9

\font\eightrm=cmr8

\def\qed{\hbox{${\vcenter{\vbox{			
   \hrule height 0.4pt\hbox{\vrule width 0.4pt height 6pt
   \kern5pt\vrule width 0.4pt}\hrule height 0.4pt}}}$}}

\begin{document}


\normalsize\textlineskip
\thispagestyle{empty}
\setcounter{page}{1}

\copyrightheading{}			

\vspace*{0.88truein}

\fpage{1}
\centerline{\bf RELATIVISTIC SELF-DUAL CHERN-SIMONS SYSTEMS:}
\vspace*{0.035truein}
\centerline{\bf  A PERSPECTIVE\footnote{To be appeared in the
proceedings of the Low Dimensional Field Theory Workshop at Telluride
(August 1996)}} 
\vspace*{0.37truein}
\centerline{\footnotesize Kimyeong Lee\footnote{Email address:
klee@phys.columbia.edu}} 
\vspace*{0.015truein}
\centerline{\footnotesize\it Physics Department, Columbia University}
\baselineskip=10pt
\centerline{\footnotesize\it New York, New York 10027, USA}
\vspace*{0.225truein}

\vspace*{0.21truein}
\abstracts{The self-dual systems are constrained and so are simpler to
understand. In recent years there have been several studies on the
self-dual Chern-Simons systems. Here I present a brief survey of works
done by my collaborators and myself. I also discuss several questions
related to these self-dual models.}{}{}


\vspace*{1pt}\textlineskip	
\section{Introduction}          
\vspace*{-0.5pt}
\noindent
In three dimensional spacetime, besides the Maxwell term, the
parity-violating  Chern-Simons term can be the kinetic part for the
gauge field.\cite{siegel,deser} The Chern-Simons Lagrangian for  an
abelian gauge field $A_\mu$ is given as
\begin{equation}
{\cal L}_{CS} =\frac{\kappa}{2}
\epsilon^{\mu\nu\rho}A_\mu\partial_\nu A_\rho, \label{cs1}
\end{equation}
and the corresponding term for a nonabelian gauge field $A^a_\mu$ is
\begin{equation}
{\cal L}_{CS}' = \frac{\kappa}{2}
\epsilon^{\mu\nu\rho}(A_\mu^a\partial_\nu A_\rho^a +
\frac{1}{3}f^{abc}A^a_\mu A^b_\nu A^c_\rho ), \label{cs2}
\end{equation}
where the coefficients $f^{abc}$ are the structure constants of the
gauge group. The Lagrangians (\ref{cs1}) and (\ref{cs2}) are invariant
under infinitesimal gauge transformations which vanish at spatial
infinity.  For quantum amplitude $\exp( i\int d^3x {\cal
L}'_{CS})$ to be invariant under large gauge transformations, the
coefficient $\kappa$ of Eq. (\ref{cs2}) should be
quantized.\cite{deser}

The abelian theory of a complex scalar field $\phi$ coupled to $A_\mu$
with the Chern-Simons kinetic term is defined by the Lagrangian
\begin{equation}
{\cal L}_1 = {\cal L}_{CS} + D_\mu\phi^\dagger D^\mu \phi - U(\phi),
\label{acsh}
\end{equation}
where $D_\mu \phi = (\partial_\mu - iA_\mu)\phi$. We will consider the
case of the pure Chern-Simons kinetic term as the Maxwell term, if
present additionally, will affect only the short distance physics.
The Gauss law of the theory (\ref{acsh}) is
\begin{equation}
\kappa F_{12} = J_0,
\label{gauss}
\end{equation}
where $J_\mu = i(D_\mu \phi^\dagger \phi - \phi^\dagger D_\mu \phi)$.
Thus total magnetic flux $\Psi = \int d^2x\, F_{12}$ is related to
total charge $Q=\int d^2x \, J_0$ by $\kappa \Psi = Q$.  The basic
excitations of the system are either charge neutral particles or
charge-flux composites.

The conserved angular momentum for the Lagrangian (\ref{acsh}) is
\begin{equation}
J = - \int d^2x \, \epsilon^{ij} x^i \left( D_0\phi^*D_j \phi +
D_j\phi^*D_0\phi \right), 
\end{equation}
whose density is gauge-invariant and localized. Under the CTP symmetry
the sign of the angular momentum does not change, and so particles and
antiparticles carry the same spin.  In the symmetric phase particles
carry nonzero spin $1/(4\pi\kappa)$. In the broken phase there are
elementary neutral scalar and vector bosons, and also charged magnetic
flux vortices whose spin is $-\pi\kappa$. Note the sign difference of
anyon spins in the symmetric and broken phases.

In two dimensional space particles of fractional spin, anyons, are
possibilities.\cite{anyon} In the symmetric phase of the
Chern-Simons-Higgs systems, anyons are represented by charge-flux
composites. The fractional statistics can be understood by considering
the orbital angular momentum of a pair of anyons or anyon-antianyon
interacting under a central force. In the system of two identical
anyons of spin $s$, the allowed orbital angular momentum is $L=2l+2s $
with an integer $l$ and so the total angular momentum is $J=L+2s =
2l+4s$. For the system of anyon-antianyon, the allowed orbital angular
momentum is $L= 2l-2s$ so that the total angular momentum is $J=L+2s=
2l$. This makes possible to creat pairs of anyon and antianyon by
vacuum fluctuations.  The fractional statistics arise when we gauge
away the flux carried by charged bosons. For the detail of anyon
physics, the readers can consult many review articles.\cite{lerda}

There are already a few reviews on the Chern-Simons Higgs
systems.\cite{khare,dunne3} This talk is a brief survey of the
relativistic Chern-Simons systems, focusing on work done by my
collaborators and myself.  In Sec.~2, I summerize the salient features
of the self-dual abelian Chern-Simons-Higgs model. In Sec.~3, the
self-dual models with nonabelian gauge symmetry are discussed. In
Sec.~4, the self-duality is generalized to the sigma and $CP(N)$
models.  In Sec.~5, some general ideas, like supersymmetry, the
correction to the Chern-Simons coefficient, and moduli space
approximation of low energy dynamics of vortices, are discussed. Here
I summerize some questions whose answer seems not known. Section~6
contains concluding remarks.

\section{ Self-dual Abelian Chern-Simons Higgs Systems}
\noindent
One of the first self-dual models found is the self-dual Maxwell-Higgs
system.\cite{bogo} The relativistic self-dual Chern-Simons-Higgs
system has been found later on.\cite{hong} The self-dual Lagrangian is
given by Eq.({\ref{acsh}) with a specific potential
\begin{eqnarray}
 U(\phi) =  \frac{1}{\kappa^2}|\phi|^2(|\phi|^2-v^2)^2.
\end{eqnarray}
The theory is renormalizable and the only dimensionful parameter is
$v^2$. When $v^2$ vanishes, there is a classical scaling symmetry
which may be broken quantum mechanically. With the help of the Gauss
law (\ref{gauss}), the energy  of the model can be expressed
as
\begin{eqnarray}
 E = \int d^2 x \left\{ |D_0\phi \pm
\frac{i}{\kappa}(|\phi|^2-v^2)\phi|^2
+ |D_1\phi \pm i D_2\phi |^2 \right\} \pm v^2 \Psi,
\end{eqnarray}
where there is no boundary contribution as we consider only finite
energy configurations. The Bogomolny bound on the energy is then
\begin{equation}
E \ge \pm \frac{v^2}{\kappa} Q.
\label{bogom}
\end{equation}
This bound is saturated by configurations satisfying the self-dual
equations,
\begin{eqnarray}
  D_0\phi \pm \frac{i}{\kappa}(|\phi|^2-v^2)\phi = 0, \,\,\,\,\,\,\,\, D_1
\phi \pm i D_2\phi =0.
\end{eqnarray}
The above equations imply that $\partial_0 |\phi|=0$ and so the field
configuration can be static in time in a given gauge choice.  Combined
with the Gauss law (\ref{gauss}), the self-dual equations can be put
to
\begin{equation}
\partial_i^2 \ln |\phi|^2 -4|\phi|^2(|\phi|^2-v^2) = 4\pi   \sum_a
\delta^2(x^i-q^i_a),
\end{equation}
where $q^i_a$'s are the positions of vortices.

The potential has two degenerate minima; the symmetric phase where
$<\phi>=0$ and the broken phase where $<\phi>=v$. As mentioned before,
there are elementary excitations in both phases and self-dual anyonic
vortices in the broken phase. In the symmetric phase there are also
self-dual anyonic nontopological solitons of unquantized magnetic
flux.\cite{jack1}

In the broken phase the self-dual configurations are determined
unquely by vortex positions.\cite{wang} While the energy is
degenerate, the angular momentum is a complicated function of vortex
positions.\cite{kim1} The statistics of anyons in the symmetric phase
is decided by the Aharonov-Bohm phase.  The statistics of anyonic
vortices in the broken phase have the contributions from both the
Aharonov-Bohm phase and a phase originated from the Magnus force.
These two phases can be combined into a single dual phase in the dual
formalism where vortices appear as charged elementary particles,
explaining the fore-mentioned sign difference of anyon
spins.\cite{kim1,mark}

The nonrelativistic limit of this self-dual model has been found and
studied.\cite{jack2} The self-dual systems with the both Maxwell and
Chern-Simons kinetic terms have been also found.\cite{clee1} This
self-dual model interpolates smoothly between the Maxwell-Higgs and
Chern-Simons-Higgs systems.

A further generalization of these self-dual models by including unform
background charge has been found.\cite{klee2} The whole structure of
this model is quite rich. Some phase appears to be infinitely
degenerate, some self-dual solitons have a negative rest mass even
though their kinetic mass is positive, in some phase there is a roton
mode among elementary excitations, some phase appears to be
inhomogeneous, implying spontaneous breaking of translation symmetry,
etc. The definition of the angular momentum becomes delicate as in the
Maxwell-Higgs case with the background charge.\cite{klee1}

\section{Self-dual Nonabelian Chern-Simons-Higgs Systems}
\noindent
The previous abelian self-dual model can be generalized to the
self-dual systems with nonabelian gauge group.\cite{klee3,dunne1} The
crucial point is to require that there exists at least a global $U(1)$
symmetry. For simplicity we consider the theory of a complex scalar
field $\phi$ in a given irreducible representation of the gauge
group. If the representation is real, we need two real scalar fields
to make the matter field complex. The generators of the symmetry group
in this representation are hermitian matrices $T^a$.  The conserved
global $U(1)$ symmetry is generated by a global phase rotation, $\phi
\rightarrow e^{i\alpha}\phi$.  The self-dual Lagrangian is then
\begin{eqnarray} 
{\cal L}_2 = {\cal L}_{CS}'
+ |D_\mu \phi|^2 -\frac{1}{\kappa^2} |T^a\phi \phi^\dagger T^a \phi
-v^2 \phi|^2,
\end{eqnarray}
where $D_\mu \phi = (\partial_\mu -iT^aA^a_\mu)\phi $. The energy
bound is given by Eq.(\ref{bogom}) with the global charge 
$Q=i(D_0\phi^+ \phi -\phi^\dagger D_0\phi)$. 

Similar self-dual models with the pure Yang-Mills kinetic term are
possible. However, there seem no interesting solitons here.  These
models may be regarded as a bosonic part of theories with an extended
supersymmetry.

Interesting vacuum and soliton structures show up when the matter
field is in adjoint representation. The vacuum expectation value of
the potential satisfies the algebraic equation
\begin{equation}
 [ [ \phi,\phi^\dagger],\phi]=v^2\phi,
\label{vacuum}
\end{equation}
where $\phi =\phi^a T^a$. This equation is the $SU(2)$ Lie algebra
with identification $J_3 = [\phi,\phi^\dagger] /v^2$ and $J_+ =
\phi/v$. This allows the detail analysis of vacuum and mass
spectrum.\cite{kao1,dunne2} The solitonic structure in the $SU(3)$
case has been studied in detail.\cite{kao1}

The nonrelativistic limit of this theory represents a theory of anyons
with nonabelian statistical phase. There are extensive work and review
of self-dual solitons in this limit.\cite{dunne4,dunne3} The dynamics
of vortices in the broken phase may involve the nonabelian
generalization of the Magnus force.

\section{Sigma and $CP(N)$  Models}
\noindent
The sigma model has been studied extensively, where self-duality has
also been explored.\cite{zak} The self-dual field configurations are
topological lumps which are characterized by the second homotopy of
the field as a mapping from two dimensional space to the internal
field space.  By gauging a part of the global symmetry of the sigma
model, one can have another self-duality. Especially a new self-dual
sigma model with the Maxwell term has been found
recently.\cite{schroers} This has been further generalized to the
models with the Chern-Simons term.\cite{ghosh,kimm1}

These models have been generalized further to the $CP(N)$ models where
the matter field is a complex vector field $z=(z_1,...,z_{N+1})$ of
unit length. The nontrivial generalization is achieved by gauging the
part of the global $SU(N+1)$ symmetry such that there exists at least
one conserved global $U(1)$ group which commutes with the gauge
group.\cite{kimm2} With the gauge symmetry generators $T^a$ and the
global symmetry generator $R$, the covariant derivative is
\begin{equation}
\nabla_\mu z =  (\partial_\mu - iT^aA^a)z - z (\bar{z}
(\partial_\mu - iT^aA^a)z ),
\end{equation}
and the Lagrangian is 
\begin{eqnarray}
{\cal L}_{CP(N)} = {\cal L}_{CS}' + |\nabla_\mu z|^2 -
\frac{1}{\kappa^2}  \biggl| \bigl(T^a z - z(\bar{z}T^az) \bigr)
\bar{z}T^a z - v \bigl(Rz - z(\bar{z}Rz)\bigr)\biggr|^2.
\end{eqnarray}
The conserved  topological current is then
$K^\mu = -i\epsilon^{\mu\nu\rho} \partial_\nu (\bar{z}(\partial_\rho 
-iA_\rho) z)$.
The global current for  $R$ is 
$J^\mu = i(\nabla^\mu \bar{z} (Rz - z(\bar{z}Rz)) - h.c.)$.
The Bogomolny bound is given by $E\ge |T|$ where $T = \int d^2x (K^0+
vJ^0/\kappa)$.

In certain limits the self-dual Chern-Simons $CP(N)$ models approach
all known self-dual Chern-Simons Higgs models, implying that the
$CP(N)$ models have all the complicated vacuum and soliton structures
as the Higgs cases, and more.  Especially with the matter in adjoint
representation, the vacuum condition is identical to Eq.(\ref{vacuum})
for an appropriate range of $v$. The structure of the self-dual
configurations is not yet fully explored.

\section{General Ideas and Questions} 
\noindent
The self-dual Chern-Simons-Higgs systems are renormalizable while the
self-dual $CP(N)$ models are not. We can use the perturbative approach
to calculate the quantum effect in the Higgs case, which has not been
fully explored even at one-loop.  When one introduces uniform
background charge, the theory is still renormalizable as far as the
dimensional counting is concerned. As the Lorentz symmetry is lost in
this case, there may be some surprises.

In the abelian case uniform external charge can be introduced and is
neutralized by the Higgs charge in the broken phase.  One can ask
whether there exists a similar configuration in nonabelian self-dual
systems. I think that it does because one can imagine a configuration
where the global $U(1)$ charge is distributed unformly, exactly like
Q-matters.\cite{coleman1}

There is also a curious homogeneous configuration whose properties are
not fully explored. The vector potential here is uniformly rotating;
$(A_1+iA_2) = c e^{iwt}$. The energy density is homogeneous and
constant parameters $c,w$ are determined by the field
equations.\cite{ezawa} We do not know whether such configuration is
classically or quantum mechanically stable. This case may be somewhat
analogous to the uniform current case in the Maxwell-Higgs case where
the current decays by nucleating vortex loops.\cite{kao2}

\newpage

\subsection{Supersymmetry}
\noindent
For every self-dual model we expect that there exits an underlying
supersymmetry.\cite{witten1} The $N=2$ supersymmetric models behind
the self-dual Chern-Simons-Higgs system have been found.\cite{clee2}
The central charge of the $N=2$ supersymmetry gives the Bogomolny
bound on the Hamiltonian. In the broken phase, the supermultiplet of
massive neutral vector bosons can have the spin structure $\pm
(1,1/2,0,-1/2)$ at most as there is only one degree of freedom
associated with vector bosons. Thus the maximal supersymmetry allowed
in this models is $N=3$.\cite{kao3} When there is uniform background
charge, the supersymmetry is not obvious at all as the mass spectrum
of a given supermultiplet does not show the
degeneracy.\cite{klee2,klee1}

In the $N=2$ or $N=3$ supersymmetric cases the Bogomolny bound is
expected to be exact because the charged sector saturating the energy
bound has a reduced representation and the supersymmetry is not
supposed to be broken. The $N=2$ supersymmetric theories needs the
infinite renormalization.\cite{leblanc,avdeev} On the other hand the
$N=3$ supersymmetric theories seem to be finite at least one
loop.\cite{kao4,kao5} It would be interesting to find out whether the
$N=3$ models are finite in all orders.

When the parameter $v$ vanishes, the classical field theory has the
scaling symmetry, which may be broken quantum mechanically by the
Coleman-Weinberg mechanism.\cite{coleman2} Indeed recently such
mechanism is shown to work here by calculating two-loop
diagrams.\cite{hosotani} The scale symmetry may be preserved quantum
mechanically for the $N=2,3$ supersymmetric models. If this is the
case, these supersymmetric theories have a quantum superconformal
symmetry.

Recently there has been a considerable progress in understanding of
the low energy nature of the self-dual Yang-Mills Higgs systems with
the $N=4$ supersymmetry in three dimensions.\cite{seiberg} Similar to
these systems, in Chern-Simons-Higgs systems magnetic monopole
instantons exist.\cite{klee4} It would be interesting if one can make
similar exact statements for the $N=2,3$ supersymmetric
Chern-Simons-Higgs systems.

Following an argument similar to that for getting $N=3$ for the
maximally supersymmetric Chern-Simons-Higgs systems, one can see the
maximal supergravity theory with massive gravitons should be $N=7$. It
would be interesting to see whether this theory, if constructed, is
finite.

\subsection{Chern-Simons Coefficient in the Broken Phase}

In the abelian Chern-Simons theories the Coleman-Hill theorem states
that the Chern-Simons coefficient does not get corrected except by the
fermion contribution at one loop when the gauge symmetry is not
spontaneously broken and there is no massless charged
particle.\cite{coleman3} The vacuum polarization by the fermion loop
renormalizes the bare Chern-Simons coefficient at the scale larger
than the fermion Compton length.  When the gauge symmetry is partially
broken, the correction to the coefficient for the unbroken gauge group
is shown to be still quantized\cite{chen}.

This theorem has been extended to the broken phase, where the `total'
Chern-Simons term is argued to be a sum of the `pure' renormalized
Chern-Simons term plus an `effective' term involving the scalar field
which looks like the Chern-Simons coefficient.\cite{khare2} One-loop
correction to the pure coefficient is quantized. That to the effective
coefficient is not quantized in general. In the self-dual abelian
Chern-Simons-Higgs system, the correction to the effective term is
however quantized.\cite{kao4} This may be true even with nonabelian
gauge symmetry with the pure Chern-Simons kinetic term.  It would also
be interesting to find out whether there is a quantum correction to
the vortex spin and if it does, whether it is related to the
correction to the coefficient.

Suppose that many family of bosons become massive fermions by a
Chern-Simons interaction and they are coupled to another gauge field.
The natural question is whether these composite fermions induce the
Chern-Simons term to another gauge field. If it does, the composite
fermions can be treated as fundamental fermions.

\subsection{Low Energy Dynamics of Vortices}
\noindent
In the broken phase of the theory considered in Sec.2, the self-dual
configurations for $n$ vortices, with gauge equivalent configurations
identified, form a finite dimensional moduli space. The natural
coordinates for this moduli space are the vortex positions
$q^i_a,a=1,,,n$. One expects the low energy dynamics of these vortices
can be described as dynamics on the moduli space.\cite{manton} There
is no potential energy for vortices as the energy is degenerate.
However there exists a term linear in velocities as the total angular
momentum depends on vortex positions.\cite{kim1} This linear term
leads to the statistical interaction between vortices and is
originated from the sum of the naive gauge interaction and the Magnus
force.\cite{kim1} The most general nonrelativistic Lagrangian for the
moduli coordinates is then
\begin{equation}
L = \frac{1}{2} T_{ab}^{ij}(q^k_c) \, \dot{q}_a^i \dot{q}_b^j +
H^i_a(q^k_c) q^i_a.
\end{equation}
One may interpret $T_{ab}^{ij}$ as the metric and $H^i_a$ as a linear
connection or a vector potential. The connection $H^i_a$ has been
obtained explicitly in terms of the self-dual
configurations.\cite{kim1} However no satisfactory answer for
$T^{ij}_{ab}$ has been found in spite of several
attempts.\cite{kim1,dziarmaga} This situation contrasts to the
self-dual Maxwell-Higgs case.\cite{samols}

When there is a uniform background charge, moduli space approximation
becomes more interesting. Again $H^i_a$ is known but $T_{ab}^{ij}$ is
not.\cite{klee1,klee2} If moduli space approximation is reasonable, a
single vortex moves a circular motion on this background due to the
Magnus force. In some range of the parameter space, the rest mass of
vortices becomes negative.  (The total energy of a pair of vortex and
antivortex is positive and so the system is stable.) I do not have any
clue for the method to calculate the kinetic mass in this case.  On
the other hand the energy difference of the Landau levels after
quantization can be larger than the rest mass of some elementary
neutral quanta in the broken phase. This contradicts the spirit of
moduli space approximation where we expect only zero modes to be
excited at low energy. This makes me to wonder whether moduli space
approximation is good at all here.

\section{Concluding Remarks}
\noindent
The relativistic self-dual Chern-Simons systems come with many
flavors. Their vacuum and soliton structures are rich and diverse.
They have been a playing ground for testing and sharpening our
understanding of quantum field theory of anyons and solitons. I have
discussed many ideas and questions related the Chern-Simons systems.

There are also many interesting topics I have not discussed at all:
the potential force between vortices away from self-duality, the
finite temperature correction to the Chern-Simons coefficient in the
symmetric phase, the theories on compact Riemann surfaces, quantum
Hall effects and boundary states, semi-local solitons, supergravity
models behind the self-dual models, etc. I believe that there are
still more surprises and insights to be discovered in this field.

\nonumsection{Acknowledgements}
I thank the organizers, Daniel Cangemi and Gerald Dunne, and many
participants of the Low Dimensional Field Theory Workshop at Telluride
(August 1996) for warm hospitality and relaxed atmosphere. This work
is supported in part by the NSF Presidential Young Investigator
program.

\nonumsection{References}

\end{document}